\documentstyle[amstex,epsf,psfig]{mn}
\newcommand{\bm}[1]{\boldsymbol{#1}}
\newcommand{\mat}[1]{\textbf{\textsf{#1}}}
\title{The use of light polarization in weak-lensing inversions}

\author[E. Audit and J.F.L. Simmons]
{Edouard Audit$^{1}$, John F.L. Simmons$^{2}$ \\
  $^{1}$ Laboratoire d'Astrophysique Extragalactique et de Cosmologie 
   -  CNRS URA 173 - \\
  Observatoire de Meudon - 5, Place Jules Jansen - 92
  195 Meudon - France\\
  $^{2}$ Department of Physics and Astronomy, University of Glasgow
  Glasgow G12 8QQ - U.K}

\date{Accepted 199-
      Received 199-
      in original form 199-}

\pubyear{1997}

\begin{document}
\onecolumn

\maketitle

\begin{abstract}
  The  measurement of  the  integrated optical  polarization of weakly
  gravitationally lensed galaxies can provide considerable constraints
  on   lens  models. The  method  outlined   depends on  fact that the
  orientation of the direction of optical polarization is not affected
  by weak gravitational lensing. The angle between the semi-major axis
  of  the imaged  galaxy   and  the direction of   integrated  optical
  polarization thus informs one   of  the distortion produced  by  the
  gravitational    lensing. Although    the  method   depends  on  the
  polarimetric measurement   of faint  galaxies, large telescopes  and
  improved  techniques should make  such measurements  possible in the
  near future.
\end{abstract}


\section{Introduction}
Gravitational  lensing  of  distant   galaxies by  foreground   galaxy
clusters provides a powerful method of determining the distribution of
matter on a large  scale.  Its strength  lies in the  fact that it  is
sensitive   to all gravitating matter, and   that it  gives the matter
distribution directly,    without  invoking  other  physics  or  other
assumptions. In this way it  is different from  other methods based on
such mechanisms as the Sunyaev-Zeldovich effect which depend on the on
the hydrodynamical  state  of the cluster (\cite{Wu96}  and references
therein).

There are  of course weaknesses  of  the method  which arise from  the
limited data available,  and the difficulty  with which these data are
obtained. With the   improvement of technique  we can  expect these
problems to be less of an obstacle in the future.

Considerable effort has     gone into inverting   images   produced by
gravitational  lensing by  clusters (\cite{Koch90},  \cite{Lup97},
\cite{Bon93}, \cite{Sei96}) to     find the   properties of    the
gravitational field and mass of the lensing cluster. One distinguishes
two regimes - that of  strong lensing, where  the mass density is near
the critical mass density for the lensing system, which produces arcs,
extreme distortions, and multiple images of galaxies, and weak lensing
where only minor distortions are observed.

In this paper we concentrate entirely on the weak lensing regime which
is appropriate  for  the outer part of  galaxy  cluster.  In the  weak
lensing regime  one is looking for  distortions of background galaxies
which are at high red-shift and very faint. Redshifts are difficult to
determine  for most imaged  galaxies,  and the determination  of their
shape subject to considerable uncertainty.  The effect of weak lensing
on an  elliptical source, as  might be  provided  by an spiral  galaxy
inclined  to  the line  of  sight,  is  to change  its  apparent size,
orientation and ellipticity.  In  principle measurement of this change
can yield  information about the gravitational  field of the lens, and
hence the matter distribution within the cluster.  In practice this is
difficult because one lacks information. Even if one assumes that both
the distance of the lens and that of the source can be determined from
redshift  analyses, this still  leaves an uncertainty in the intrinsic
size, ellipticity   and orientation   of  the source   galaxy.   These
uncertainties     have led researchers  to   treat  the   problem as a
statistical  one,  where the  background galaxies    are treated as  a
statistical  ensemble without  any  attempt to  use  information about
their     intrinsic   properties.   (\cite{Bon93},   \cite{Kai95},
\cite{Sei96})

If  in  some way one's ignorance   of  the source parameters  could be
reduced the  lens parameters could  be better determined.  Measurement
of the apparent brightness and   redshift are obvious ways to  improve
inferences, but  are difficult to achieve  given the faintness  of the
images.  Another interesting property that  has yet to be exploited is
the optical polarization of the source galaxy. Measurement of galactic
polarization of high redshift  galaxies would of course  be difficult,
given the faintness of the objects, although this has been carried out
for a number of radio galaxies in  both radio \cite{Gab92} and optical
bands \cite{Tad92}.  Potentially  it could yield valuable information. 
(Indeed, in  the  case of radio  observations these  measurements have
been  carried  out with considerable   accuracy with  a  resolution of
milliarcseconds.) The reason is  the following. Light  polarization is
not affected,  at least if    the lensing  object has  small   angular
momentum (in relativistic units) \cite{Sch92} by gravitational lensing
in the sense  that the direction of  polarization will not be altered. 
For a typical spiral galaxy containing  free electrons and dust in the
galactic  disc  the  integrated   optical  polarization  arising  from
scattered starlight could be as high as 2 percent and one would expect
it to be in  a direction perpendicular to  the disc  \cite{Bianchi96}. 
Even if other mechanisms such as  dichroic extinction are at play, the
reasonable assumption of large  scale axial symmetry requires that the
integrated polarization should be either in the direction of the major
axis of   the ellipse  or   perpendicular to  it. Thus  the integrated
polarization could in principle be used  to fix the orientation of the
source  galaxy, modulo 90  degrees. 

The   degree of polarization  will  obviously depend  on  the angle of
inclination of the source galaxy to  the line of  sight and the number
of scattering particles in the galactic disc.  If we assume Thomson or
Rayleigh scattering and axial symmetry in  the distribution of sources
and   scatterers, one  can   show   \cite{SA97}  that   the degree  of
polarization for optically thin galaxies  is proportional to $ \sin ^2
i $ and the optical depth along the  galactic radius, where $i$ is the
angle of inclination of the axis of symmetry of the galaxy to the line
of sight.

One signature of a lensed galaxy would be an observed deviation of the
direction of polarization  from  the perpendicular to  the  semi-major
axis of  the  image ellipse.   In  principle  it is  possible for  the
direction of  integrated polarization to  be  in the  direction of the
semi-major axis of the ellipse, owing to other mechanisms as discussed
by  several  authors    (\cite{Draperetal95},   \cite{Scarrott90},
\cite{Wood97} and \cite{Woodetal97}).  Small scale deviations from
rotational   symmetry    produced  by  clumpiness  in  the  scattering
distribution,       as   discussed by   \cite{WittandGordon96},   or
inhomogeneities in  the  galactic  magnetic  field  in the    case  of
dichroism,  should not  greatly   affect the direction   of integrated
polarized flux although the degree of integrated polarization could be
reduced.

In  the ideal  situation  the degree    of polarization should   yield
information about the inclination of the source galaxy  to the line of
sight.  It would in fact, however,  be difficult to accurately measure
the  degree  of polarization, or  to  unambiguously infer from  it the
inclination,  and  hence the  ellipticity  on the   sky  of the spiral
galaxy. (High inclination gives rise to high polarization.)

In   this paper we    investigate how measurement   of galactic  light
polarization can be used  to sharpen and  improve the determination of
the  lens parameters.  In section (\ref{simplemodel})  we give a rough
estimate of the degree of polarization produced in a spiral galaxy due
to dust and electron scattering. Our main purpose in  doing this is to
obtain a ball-park value for the degree of polarization. More accurate
modelling    has  been  carried  out  for     the  optically thin case
\cite{SA97},  and by   Monte Carlo  treatment   of radiative  transfer
(\cite{Wood97},  \cite{Woodetal97} and \cite{Bianchi96}) for the
optically  thick case,   but with the   view to  obtaining  a resolved
polarization map rather  than the integrated polarization.  In section
(\ref{inv})  we discuss  how  information  on the  inclination of  the
source galaxy together    with  measurements of  the ellipticity   and
orientation of the  images can be used  to infer the parameters of the
lens, and in section (\ref{comp}) we compare the results obtained when
this polarimetric  information is available with the   case when it is
not,  in  order to assess   the usefulness of   the  method.  The last
section contains a short discussion of the feasibility of carrying out
such a polarimetric measurement and our conclusions.

\section{Galactic polarization}\label{simplemodel}

Measurement  of   optical polarization of   spiral  galaxies  has been
carried  out   for a    few   nearby  galaxies   (\cite{Scarrott90},
\cite{Nei90}).    Various mechanisms  could  produce  this   optical
polarization, the most obvious being scattering by dust and electrons.
Most  recent  studies however  have largely   been  concerned with the
detailed  structure  of the  galaxies  and the  possible  existence of
appreciable magnetic fields.  Accordingly they have aimed at providing
a resolved map  of polarization in  optical and infrared, which  could
give some information not  only  about the distribution of  scattering
grains and electrons, but also  about the presence of magnetic  fields
and of dust   lanes near the  galactic nucleus.   Dust grains, assumed
oriented  by the magnetic field  of  the galaxy, preferentially absorb
light  along their longer axis, and   hence could produce polarization
perpendicular   to this axis of  the  grain (dichroism).  Such a model
\cite{Woodetal97} has been partly succesful in explaining the observed
pattern   for  several  high  inclination  spiral   galaxies where the
extinction  would  be most  pronounced.  In   such cases,  near to the
nucleus the polarization is parallel  to the galactic plane.  However,
that  is   not   necessarily the   case   for   other  spiral galaxies
\cite{Scarrott96}.

In this paper we are interested only  in the polarized flux integrated
over the whole galaxy, which should in principle give an indication of
the orientation of  the galaxy.  If the  spiral galaxy has an axis  of
symmetry,  so  that its  isophotes are   ellipses, we should,   on the
grounds of symmetry, expect the  linear polarization flux to be either
perpendicular the  major axis  of the  ellipse, or possibly,   if such
effects as dichroism were   important, along the minor axis.  Although
the small scale deviations from axial symmetry  have been observed for
spiral galaxies at  high inclinations, we  should expect most of these
effects to  be largely cancelled out  in the integrated flux, which we
would expect   to be either  perpendicular  to the  major  axis of the
ellipse or along the major  axis. The apparent rotation brought  about
by weak lensing would not usually be more than a  few tens of degrees,
and thus the determination of the direction of integrated polarization
'modulo' 90 degrees would be sufficient  to provide pretty unambiguous
information about the orientation of the source galaxy.

In the case of spiral galaxies,  one would expect optical polarization
to  arise as a  result of  Thomson, Rayleigh  and  dust  scattering of
starlight primarily in the galactic  disc of  the spiral galaxy,  with
minor    contributions  from  the   halo. For   Thomson   and Rayleigh
scattering,  light scattered  through  90  degrees  will  be  entirely
polarized in a direction perpendicular to the  scattering plane.  (For
other  scattering mechanisms we would   expect the polarization to  be
either in the  plane or perpendicular.)  Thus  if Thomson or  Rayleigh
scattering were    dominant, a galaxy inclined to    the line of sight
should display polarization,  and one would  expect in  the case of  a
rotationally symmetric disc that  the direction of polarization to  be
along the  minor  axis of the  ellipse  \cite{Bianchi96},  although in
exceptional circumstances it could be along the major axis.

\subsection{A simple model for the integrated polarization}

\begin{figure}
\psfig{file=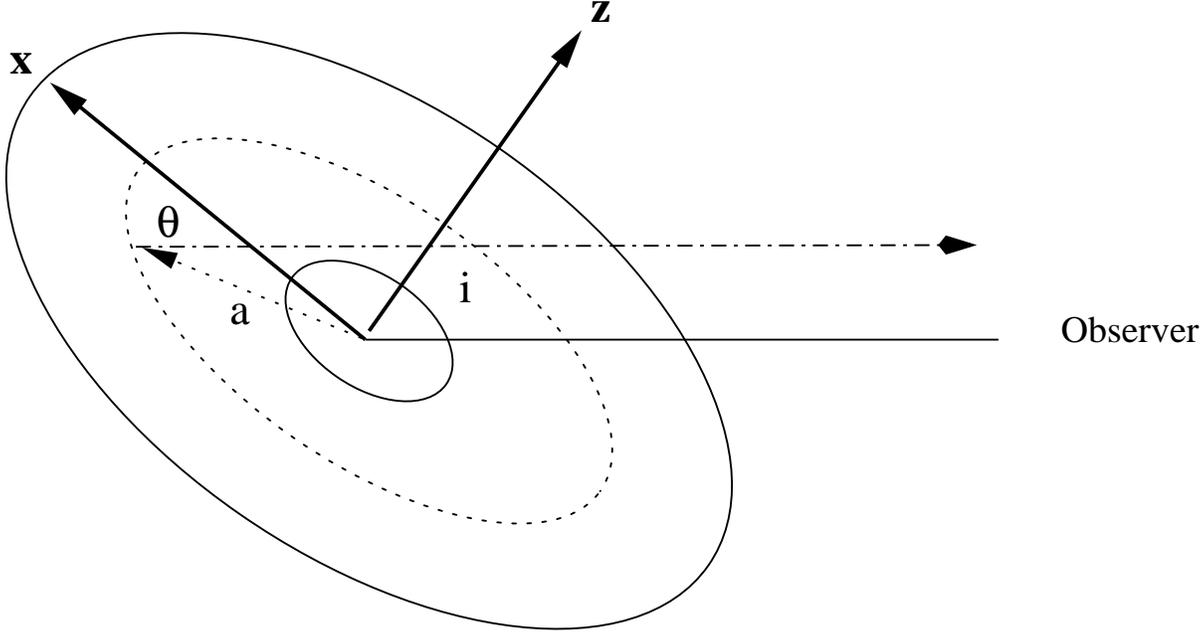,width=16.cm}
\caption{Simple scattering geometry. The galaxy is represented by a
  annulus, with inner  radius $R_1$ and outer  radius $R_2$, with  the
  source at the centre.   The angle between  the line of sight and the
  axis  of symmetry ({\bf X})  is  denoted $i$.   The ray scattered by
  element at ${\bf a}$ is indicated by dot-dashed line. }
\label{model_simple}
\end{figure} 

One  can show that  under fairly  general   assumptions the degree  of
polarization produced by Thomson or Rayleigh scattering, in the single
scattering regime  depends  on $\sin  ^2 i$  and the  total  number of
scattering particles  \cite{SA97}.  In  this  section we  shall give a
very simplified  derivation of   this result  to  obtain an  order  of
magnitude for the galactic polarization.

Consider a completely flat galactic disc with the  light source at the
centre (see Fig. \ref{model_simple} ). Take  the surface density of
electrons in the disc  to be ${\cal N}_e$  and  the luminosity  of the
galaxy to be $L$.   The axis of symmetry of  the galaxy, which we have
taken to be the z-axis, is at inclination $i$ to the line of sight. We
take the x-axis to be in the plane of symmetry.

The flux  arriving at the scattering element  at ${\bf a}$ is given by
$L/4\pi a^2$.   Thus the energy  scattered per steradian per unit time
into the line of sight by this element is simply
$$
dF=\frac{L\, \sigma}{4\pi   a^2}  {\cal N}_e  (a)  a \,   d\theta \,
da\frac{3}{16\pi} (1+\cos^{2}\chi)
$$
\noindent
where $\theta $ is the angle between ${\bf a}$ and the x-axis,  $\chi$ is the scattering angle and $\sigma$ the total scattering
cross section.  We obtain similar  expressions for the polarized flux.
In terms of the Stokes parameters referred to  the scattering plane we
have

$$
dF_Q=\frac{L\, \sigma}{4\pi  a^2} {\cal N}_e (a) a \, d\theta \,
da\frac{3}{16\pi}
\sin^2\chi
$$

and

$$dF_U=0 .$$

To obtain the total  scattered energy per steradian  per unit  time we
integrate over the disc to obtain

$$
F=\int\int \frac{L \, \sigma }{4\pi  a^2} {\cal N}_e (a) a \frac{3}{16\pi}
(1+\cos^2\chi)  d\theta da
.$$

Using the properties of the Stokes parameters under rotations, total
polarized energy $F_Q$ expressed in the observers frame (with the
polarimeter aligned with the axis of symmetry) is given by

$$
F_Q=\int\int \frac{L}{4\pi  a^2} {\cal N}_e (a) a \frac{3}{16\pi}
\sin^2\chi\,  \cos 2\phi \, d\theta da
$$

and

$$ F_U=\int\int \frac{L \, \sigma}{4\pi  a^2} {\cal N}_e (a) a \frac{3}{16\pi}
\sin^2\chi  \sin 2\phi \, d\theta da \, ,$$

\noindent
where  $\phi$ is the angle between  the scattering plane and the plane
defined by the line of sight  and the axis of symmetry  of the galaxy.
The scattering  angle $\chi$ and the angle  $\phi$ may be expressed in
terms of $\theta$ and $i$

$$\cos \chi =-\cos \theta \sin i $$

and

$$\cos\phi =\frac{\cos\theta \cos i}{\sin\chi}  .$$
\noindent
Substitution of these into the integral expressions yields after a bit
of reduction

$$
F_Q   =    \frac{L \, \sigma }{4\pi}\frac{3\pi}{8}\sin^2i  \int
\frac{{\cal N}_e(a)}{a}da
$$
and 
$$F_U=0\, ,$$ signifying  that the polarization  lies along the axis
of symmetry (because $F_Q$  is always greater or equal  to zero).  We
shall ignore the unpolarized scattered  power as this will be  swamped
by the direct  light from  the galaxy. The  degree of  polarization is
simply  $F_Q /  F_{\mbox{direct}}$  where  $F_{\mbox{direct}}$ is  the
direct  power  per   steradian,  $L /  4\pi $.    Thus   the degree of
polarization is

$$ p =\frac{3\pi\, \sigma}{8}\sin^2i \, \int
\frac{{\cal N}_e(a)}{a} da \, .$$
The degree  of polarization depends  on $\sin^2 i$.  
There   are obvious problems in    evaluating  this integral for  most
density  distributions because of  the  possible singularity at  $a=0$
which arises  from the two  dimensional  treatment of  the problem. To
avoid such difficulties we take the surface  density to be zero within
a certain radius of  the galactic centre.  Again  since we are  really
only  interested  in orders of  magnitude  here,  we take  the surface
number density  to be  constant  within an  annulus with  inner radius
$R_1$ and  outer  radius $R_2$.  We   then  obtain for the   degree of
polarization

$$
p= \frac{3\pi {\cal N}_e \sigma }{8}\sin^2i
\, \ln \frac{R_2}{R_1} .
$$

If we denote the optical depth through the galactic disc by $\tau$ and
assume a geometric thickness for  the  galactic disc of $\Delta$  then
the degree of polarization may be written
$$p= \frac{3 \tau \pi}{8}\sin^2i \, \frac{\Delta}{R_2-R_1}ln
\frac{R_2}{R_1} .$$

Typically we can expect the ratio of the  thickness of the disc to its
radius to be 0.1.  so that the degree of polarization is approximately
$p=0.1 \tau \sin^2i$ where we take the value of $ln {R_2}/{R_1}$ to be
of the order of 1. Thus with an optical depth in the range 0.1 to 1 we
obtain  a  maximum polarization of  one  to  ten  percent. The  actual
polarization depends of course  also on the  inclination of the galaxy
through the $\sin^2 i$ factor.

This simple model seems to give  the right order  of magnitude for the
integrated polarization  when compared  to  more sophisticated  models
\cite{Bianchi96}.

\section{The lens equation and weak lensing}\label{inv}
\begin{figure}
\psfig{file=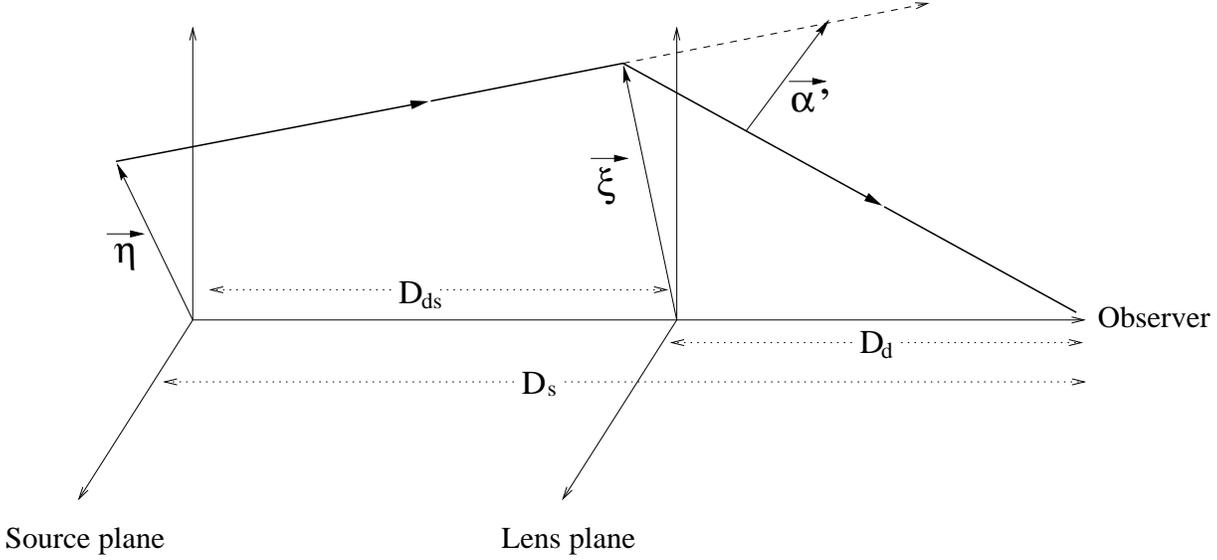,width=16.cm}
\caption{Simplified picture of the lens situation.}
\label{banc-opt}
\end{figure}
A simplified picture of the lensing situation is represented in figure
\ref{banc-opt}, which  serves  to define the quantities  entering into
our equations. The lens equation relates  the position, ${\bm {\eta}}$
of the  source  galaxy  (in the  source   plane which is  a   distance
$D_{\mbox{s}}$ from the observer) to the position, ${\bm \xi}$, of the
image galaxy in the lens plane, which  is a distance $D_{\mbox{d}}$ to
the observer.  (Distance  here   means the  angular  size distance).   
Elementary geometrical arguments yield
\begin{equation}
\bm{\eta}=\frac{D_{\mbox{s}}}{D_{\mbox{d}}}\bm{\xi}-
D_{\mbox{ds}}\bm{\alpha '}
\end{equation}
where ${\bm \alpha '}$ is the vector angular deflection
produced by the
  lens. If we use angular coordinates instead this equation may be
  written
\begin{equation}
{\bm y}={\bm x}- {\bm \alpha}\, ,
\label {lens_equation}
\end{equation}
where $ {\bm  \alpha}$ is defined to be  ${\bm  \alpha ' }
D_{\mbox{ ds} }/ D_{\mbox{ s}}$. The deflection angle is determined by
the matter distribution in the  lens plane. With the usual assumptions
(see Schneider et al. 1992) we may write
\begin{equation}
{\bm \alpha} = \nabla \Psi
\end{equation}
where
\begin{equation}
\Psi({\bm x}) = \frac{1}{\pi}\int d^2 x' \kappa '  ({\bm x '})\ln |x-x'|
\end{equation}
and     $\kappa ' =    \Sigma    /   \Sigma_{\mbox{  crit}}$,    and
$\Sigma_{\mbox{crit}}=c^2            D_{\mbox{s}}/4\pi               G
D_{\!\mbox{d}}D_{\!\mbox{ds}}$,  where $\Sigma$  is the projected mass
density.  It follows that

$$\nabla^2 \Psi = 2\kappa ' \, .$$

The amplification matrix, $\mat{A}$, is then defined as the inverse of the
jacobian matrix $(\partial y^{i}/\partial x^{j})$, and so yields

\begin{equation}
\mat{A}^{-1}= \left(
        \begin{array}{cc}
         1-\Psi_{11} &  \Psi_{12} \\
         \Psi_{12}   & 1-\Psi_{22}
        \end{array}
      \right) \, .
\end{equation}

It follows  from  a linear  expansion  of equation \ref{lens_equation}
that  in  the  weak regime, an   elliptical  source  (i.e.  elliptical
isophotes) will    be    transformed   into  an     elliptical   image
\cite{Koch90}.  The relation  between    the  two depends   on   the
properties    of the  lens,    and   is  given   below   by  equations
(\ref{s1}-\ref{s4}) in  terms of  the of  the  characteristics  of the
source and  image ellipses,   and the  lens.  Following Kochanek,   we
characterise the source ellipse by the triplet ${\cal S}=(\lambda_{s},
\Delta\lambda_{s}  ,\alpha_{s})$, the image by ${\cal I}=(\lambda_{i},
\Delta\lambda_{i}, \alpha_{i})$ ($\alpha$ gives the orientation of the
ellipse      and      $(     \lambda+\Delta\lambda)^{1/2}$   (resp.
$(\lambda-\Delta\lambda)^{1/2}$) its major (resp.  minor) axis length)
and  the lense by  ${\cal  L}=(\kappa,  \gamma, \theta)$.  Effectively
these provide

\begin{eqnarray}
\label{s1}
  \Delta\lambda_{s} S_{s} &=&
(\kappa^{2}-\gamma^{2})S_{i} \Delta\lambda_{i}\\
\label{s2}
\lambda_{s} &=& \lambda_{i} (\kappa^{2}+\gamma^{2}) + 2
\Delta\lambda_{i} \kappa \gamma C_{i}\\
\label{s3}
\lambda_{s} \pm \Delta\lambda_{s} C_{s}
&=&(\lambda_{i} \pm \Delta\lambda_{i}C_{i}
)(\kappa \pm \gamma )^{2}\\
\label{s4}
 \Delta\lambda_{s} C_{s}&=& 2 \kappa \gamma
 \lambda_{i} + \Delta\lambda_{i}C_{i}
(\kappa^{2}+\gamma^{2})
\end{eqnarray}
where    we have   defined:    $C_{x}=\cos 2(\alpha_{x}-\theta)$   and
$S_{x}=\sin 2(\alpha_{x}-\theta)$.   There are $9$  unknown parameters
and three independent  equations. Thus in principle  any one  of $\cal
S$, $\cal  I$,or $\cal L$  can be determined from   a knowledge of the
other two.   In particular the   source parameters, $\cal  S$,  can be
determined if one knows  the  image, $\cal  I$, and lens   parameters,
$\cal L$.

In what follows  we  shall also  use the parameter  $e$ describing the
source and image  ellipticity    defined in terms  of   $\lambda$  and
$\Delta\lambda$ as
$$
e=1-\sqrt{\frac{\lambda-\Delta\lambda}{\lambda+\Delta\lambda}}
\, .$$

It is  interesting to note, before  making any attempt to  solve these
equations, that  they have a number of  scaling laws which should make
their  solution easier.   Thus   for a  given lens, if  $(\lambda_{i},
\Delta\lambda_{i},   \alpha_{i},     \lambda_{s},   \Delta\lambda_{s},
\alpha_{s}   )$   is   a  solution,     then  so  is   $(x\lambda_{i},
x\Delta\lambda_{i}, \alpha_{i},   x\lambda_{s},    x\Delta\lambda_{s},
\alpha_{s} )$  where $x$ is a  constant.  Moreover, for a given image,
if $(\lambda_{s},   \Delta\lambda_{s}, \alpha_{s},    \kappa,  \gamma,
\theta)$   is   a    solution, then  so     also is  $(y^2\lambda_{s},
y^2\Delta\lambda_{s},   \alpha_{s}, y\kappa,  y\gamma, \theta)$.  This
means that  one can arbitrarily take $\lambda_{i}$  and $\kappa$ to be
$1$.         Dividing   the    equations    (\ref{s1})-(\ref{s4})   by
$\kappa^2\lambda_{i}$ ($x=1/\Delta\lambda_{i}$ and  $y=1/\kappa$)  one
obtains:

\begin{eqnarray}
\label{sr1}
\Delta l_{s} S_{s} &=& (1-\Gamma^2)S_{i} \Delta l_{i}\\
\label{sr2}
l_{s} &=& (1+\Gamma^2) + 2 \Delta l_{i} \Gamma C_{i}\\
\label{sr3}
l_{s} \pm \Delta l_{s} C_{s} &=&(1 \pm \Delta l_{i}C_{i})(1 \pm \Gamma )^{2}\\
\label{sr4}
 \Delta l_{s} C_{s}&=& 2 \Gamma + \Delta l_{i}C_{i} (1+\Gamma^2)
\end{eqnarray}
where        $\Delta         l_{i}=    \Delta\lambda_{i}/\lambda_{i}$,
$l_{s}=\lambda_{s}/(\kappa^2\lambda_{i})$,    $\Delta         l_s    =
\Delta\lambda_{s}/(\kappa^2\lambda_{i})$, and $\Gamma=\gamma/\kappa$.

\section{Inferring the lens parameters}\label{comp}

The determination of the lens parameters in the weak lensing regime is
usually  treated  as a    statistical  problem.   In the absence    of
additional information about  the source galaxies, in particular about
the orientation of their  axis  of symmetry  (if they have  one), most
analyses to date have attempted to  solve the problem of inferring the
lens  parameters  by  matching the distribution  of  source parameters
obtained from the data by the lens model, with an assumed distribution
of  source galaxy    parameters   e.g.  a   uniform  distribution   of
orientations of source  galaxies.  One can  obtain the best  fit model
parameters by  standard routines such as  $\chi^2$- fitting. This sort
of approach  has  been  adopted by   ,  for example,  \cite{Bon93}.  
Although very promising,  the   measurements are  not good  enough  to
provide accurate determination of the lens parameters.  The techniques
developed  to reconstruct the    cluster  potential from  the   lensed
galaxies are now extremly sophisticated and they treat with great care
the observational    and   statistical uncertainties  (\cite{Kai95},
\cite{Sch96}, \cite{Koch90}). We do not intend in this paper to go
into  such   details. Our point  is  just   to show  that polarimetric
measurements  put strong constraints      on the  cluster    potential
reconstuction.

We have already argued that information about the source galaxies, and
in particular  about  their orientation   on the sky,  which could  be
obtained from measurement of their optical polarization, would lead to
more  accurate   determinations.   Polarization  measurements    would
undoubtedly  be difficult for such  faint galaxies.  However, even the
measurement  of  the polarization for a   small subset  of  the galaxy
sample can considerably improve   the accuracy on the lens  parameters
determination. In the following section we illustrate this numerically
for simple lens models.

To assess what can be obtained via  polarization measurements, we have
carried out  a  simple simulation.  First  we decide on  a lens model,
which  fixes   the   values  of   $\kappa(x,y)$,   $\gamma(x,y)$   and
$\theta(x,y)$ in the lens plane. We then randomly generate a sample of
source  galaxies  from    a  given distribution  with   the  following
properties. We take the distribution of the  symmetry axis to the line
of sight  to  be isotropic, which  implies  a uniform  distribution in
orientation on the sky, and uniform distribution in ellipticity (if we
assume the galactic discs  to be circular). Since  we are going to use
the   system  (\ref{sr1})-(\ref{sr4}) which  contains only ``reduced''
variables, we do not  need to know  the brightness of the  galaxies at
this stage.These galaxies are  then  lensed by a  given lens  model to
give a distribution of images. We then attempt to reestablish the lens
parameters from   the sample of  images.  We shall  see below that the
effectiveness of the   retrieval depends crucially on  the assumptions
about the model, and whether or not we use potential information about
the  orientation of the  source  galaxies,  which  we argue, could  be
furnished by polarimetric data.

\subsection{A spherical lens model}
\begin{figure}
\psfig{file=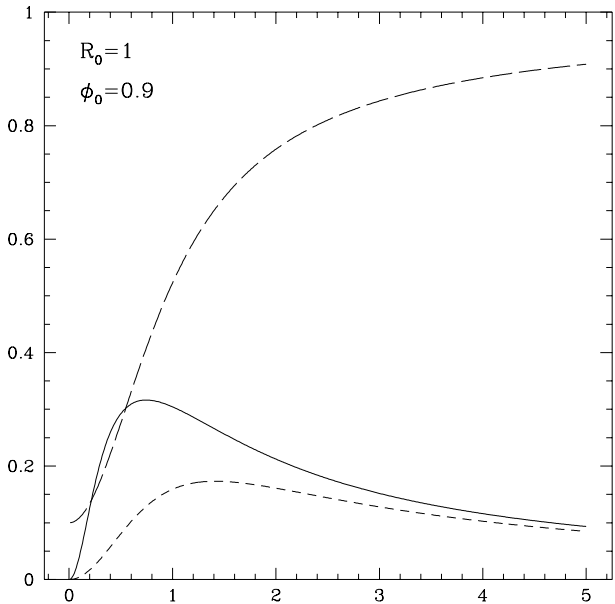,height=9.cm,width=9.cm}
\caption{Lens parameters for the isothermal sphere. $\kappa$
  long-dashed line, $\gamma$ short-dashed line, $\Gamma=\gamma/\kappa$
  full-line.}
\label{sphere_iso}
\end{figure}

To illustrate our argument  we take an  isothermal sphere with a  core
radius as  our lens model  \cite{Sch92}.  The lens parameters are then
given by
\begin{eqnarray}
\kappa& = \frac{1}{2}\Phi_{0} \left( d^{-3/2}+d^{-1/2} \right)\\
\gamma& = \frac{1}{2}\Phi_{0} \left( d^{-3/2}-d^{-1/2} \right)
\end{eqnarray}
where $d=(1+(r/r_0)^{2})$ and $r$ is the distance  to the lens
centre. In addition to  the values of  $r_0$, and $\Phi_0$, the
coordinates of the centre of symmetry, introduce  two  further
parameters.  The centre  of   the  lens  can be inferred in various
ways (see for example \cite{Koch90})  and we shall assume in the
rest of this article that it is  already known. We have used a lens
with $r_0=1$ and $\Phi_0=0.9$ (Fig.\ref{sphere_iso}).

\begin{figure}
\psfig{file=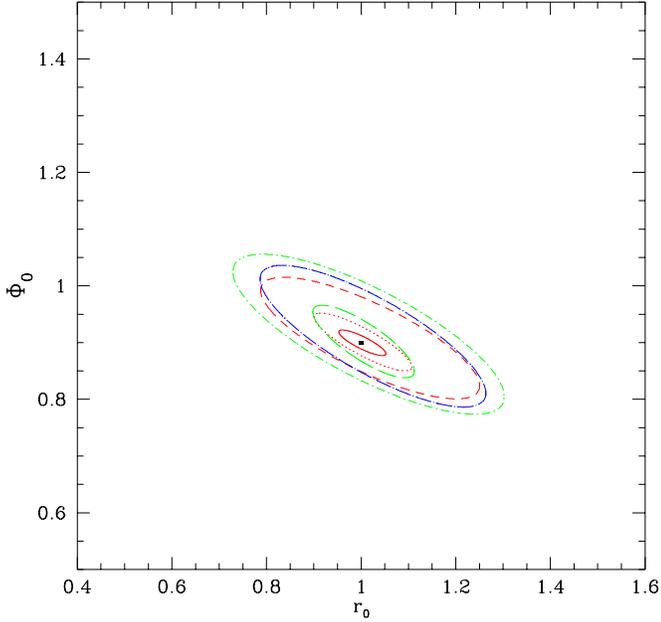,height=9.cm,width=9.cm}
\caption{This figure shows the ellipses of the $(r_0, \Phi_0)$ plane
  which contain $70\%$ of the reconstructed  points. the solid line is
  for ($N=50,    \epsilon=2\%$),   the   dotted     for   ($N=50,
  \epsilon=5\%$), the small dashed   for ($N=50,  \epsilon=10\%$),
  the long-dashed   for ($N=20, \epsilon=2\%$),   the dotted-small
  dashed for ($N=20, \epsilon=5\%$) and the dotted-long dashed one
  for ($N=10, \epsilon=2\%$).}
\label{pol70}
\end{figure}

\begin{figure}
\psfig{file=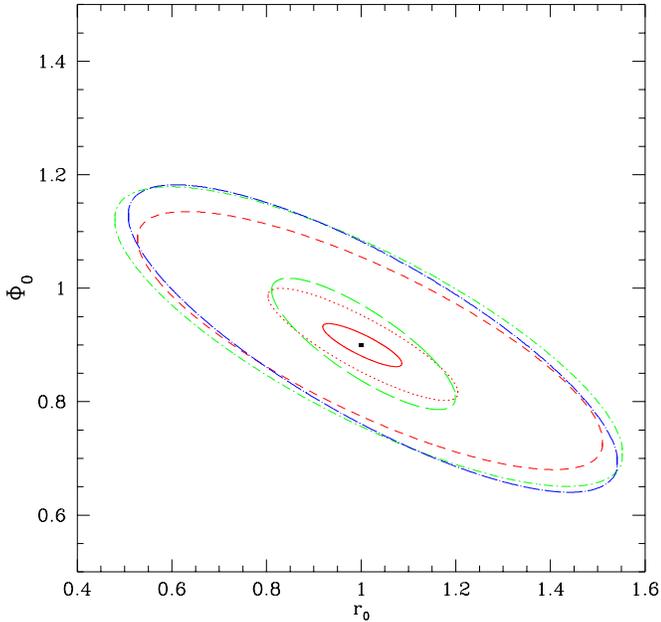,height=9.cm,width=9.cm}
\caption{This figure is the same as  Fig. \ref{pol70} but with  the
  ellipse containing $90\%$ of the points.}
\label{pol90}
\end{figure}

If we assume values for the pair of parameters $(r_0,\Phi_0)$, and the
image parameters are assumed to  have been determined exactly, then it
is possible to reconstruct the parameters for the source. If we assume
also that the  only information that we have  about the source is from
the measurement   of  the polarization,  which  provides   us with the
orientation of    the  source galaxies   with   a  standard  error  of
$\epsilon$, then we can  find  the lens parameters  that best  fit the
observed polarization  direction. The  natural way to  do  this is  to
carry out  a chi-squared minimisation,  i.e. we minimise the following
expression over the lens parameters.

\begin{equation}
\label{chi}
\chi_1(r_0,\Phi_0)=\sum_{i=1}^{N}
     \frac{(\alpha_{s}^{r}(r_0,\Phi_0)-\alpha_{s}^{m})^{2}}{\epsilon^{2}}\, .
\end{equation}
The summation  in equation (\ref{chi}) is  carried out over the set of
galaxies whose polarization  has been determined.  $\alpha_{s}^{r}$ is
the direction which has been reconstructed  from the parameters of the
lens model, and $\alpha_{s}^{m}$   is the measured direction. We  have
reconstructed  lens  parameters  for the  cases  when  the   number of
galaxies, $N$, for  which the polarization  has been measured is given
by $N=50$, $10$ and $5$. For each of these cases we consider the error
on  the polarization direction, $\epsilon$, to  be given by  2, 5, and
10\% .In order to get an idea of the confidence one should have in the
values of $r_0$  and $\Phi_0$ obtained in  this way, we have for, each
pair of values $(N,\epsilon)$,  generated  $500$ different samples  of
galaxies, and from these used  the chi-square estimator of the  values
of  $(r_0,\Phi_0)$. Figures  \ref{pol70}  and   \ref{pol90}  show
ellipses  in the $(r_0,\Phi_0)$ plane  which contain $70\%$ and $90\%$
of the reconstructed values.

\begin{figure}
\psfig{file=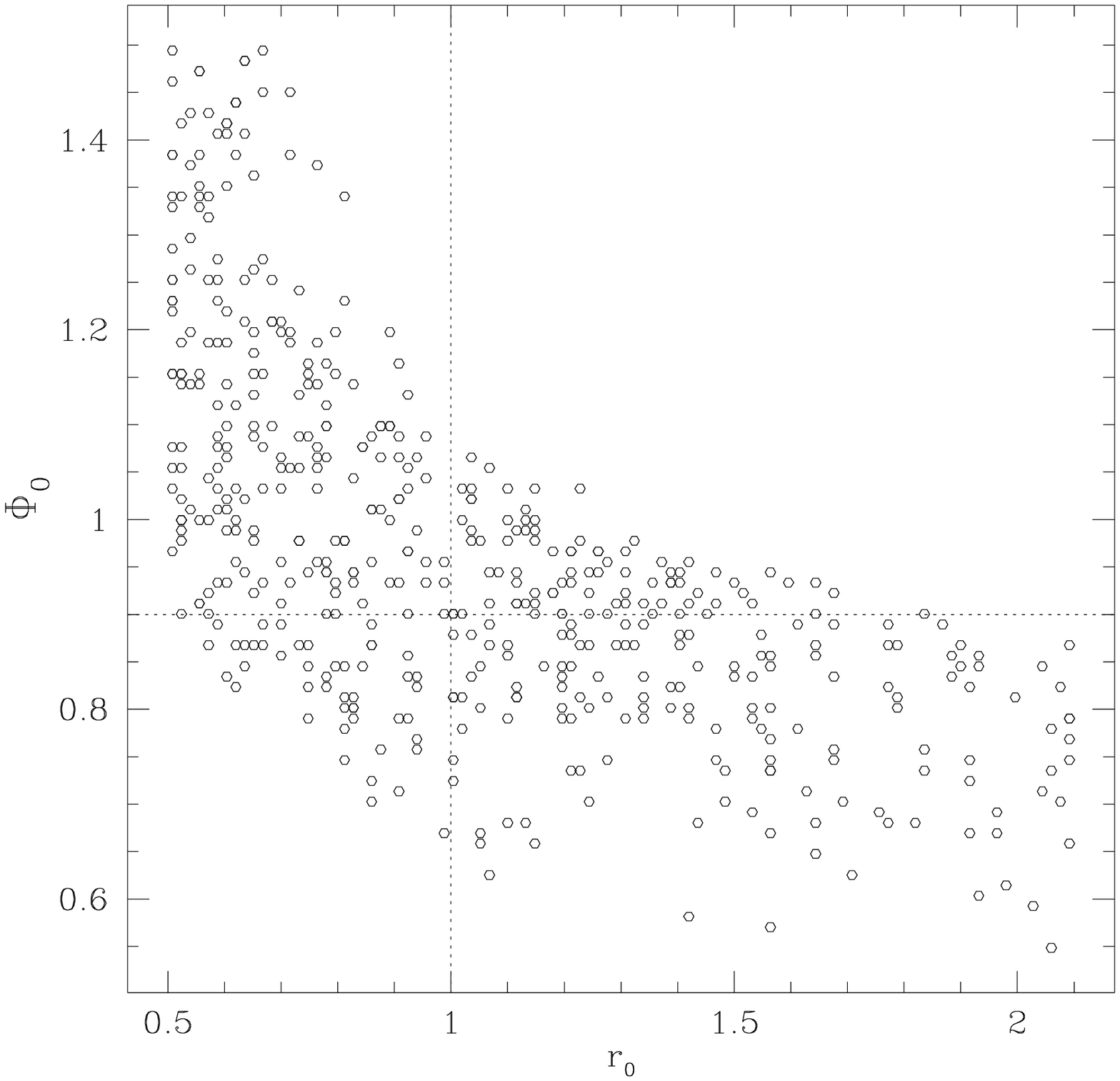,height=9.cm,width=9.cm}
\caption{Each point in this figure gives the couple $(r_0, \Phi_0)$
  reconstructed  from a sample  of  $500$ galaxies  without using  the
  information given by the polarization. (There  are no points outside
  this square because during  the minimisation process, we limited the
  variation of $r_0$ to  $[0.5, 2.1]$ and  that of $\Phi_0$ to  $[0.4,
  1.5]$ .)}
\label{recon_dis}
\end{figure}

In  order to  quantify the  information that   the  measurement of the
polarization provides,  we have also carried  out  a reconstruction of
the lens (that is determined $r_0$ and $\Phi_0$) assuming no knowledge
of the orientation of  the source galaxies. We  did this by  comparing
the distribution of source parameters inferred from an assumed pair of
lens  parameters and sample of image   galaxies with a distribution of
source galaxies that  is assumed  to  be  uniform  in orientation  and
ellipticity. We  compared the  joint distributions in  ellipticity and
orientation by chi-squared fitting  of the two distributions. First we
divide the $\alpha_s$,  $e_s$ plane  ($\alpha_s  \in  [0,\pi]$ is  the
source orientation,   and $e_s \in [0,1]$   is its ellipticity  ) into
$N_c$ bins. We then minimise
\begin{equation}
\chi_2(r_0,\Phi_0)=\sum_{i=1}^{Nc}
     \frac{(n_{r}^{i}(r_0,\Phi_0)-n_{ex}^{i})^{2}}{n_{ex}^{i}}
\end{equation}
with respect to  ($r_0$,$\Phi_0$).  $n_{r}^{i}$ is the observed number
of galaxies  in bin $i$ and $n_{ex}^{i}$  the predicted number for any
assumed pair of lens parameters. We have binned the data in such a way
that $n_{ex}^{i}=20$.  This   procedure was  carried out for   several
hundred  samples of  $500$ galaxies.  The reconstructed $(r_0,\Phi_0)$
are  plotted on figure (\ref{recon_dis}). We  can see that  there is a
great dispersion in the reconstructed data. This  arises from the fact
that the statistic of  the reconstructed sources is fairly insensitive
to  the lens parameters. Therefore the  above $\chi^2$  is small for a
wide range of parameters over which the minimum can be found.

Although this method of    finding  confidence regions for   the  lens
parameters  in the  absence of   polarization information  may not  be
optimal, it does however given  some idea of  the advantages of  using
polarimetric data (see Fig. 2).

\subsection{An arbitrary spherical lens}

As we have already mentioned, if one  does not use the polarization to
determine the orientation  of  the source galaxy, the  only measurable
quantities are  the image  parameters.   We are  thus left   with five
unknown parameters $(l_s, \Delta l_s, \alpha_{s}, \Gamma, \theta)$ but
still have  only $3$ equations,  and so it is  impossible to solve the
system.

If we assume   only that the  lens  is spherically symmetric,  and, as
before,   that the centre of   the lens is    known, so that the angle
$\theta$  is determined, and that  the polarization has been measured,
then the system of  equations (\ref{sr1}),(\ref{sr2}) and  (\ref{sr4})
will contain three  unknowns $(l_s, \Delta l_s,  \Gamma)$.  It is thus
possible   to  deduce  the  remaining  source  parameters  as  well as
$\Gamma$.  From equations (\ref{sr1})  and  (\ref{sr4}) one can   show
that $\Gamma$ is given by

\begin{equation}
\label{solr}
(C_i S_s + S_i C_s)\Delta l_{i} \Gamma^{2} + 2 S_s \Gamma
+( C_i S_s - S_i C_s)\Delta l_{i} = 0 \, .
\end{equation}

\begin{figure}
  \psfig{file=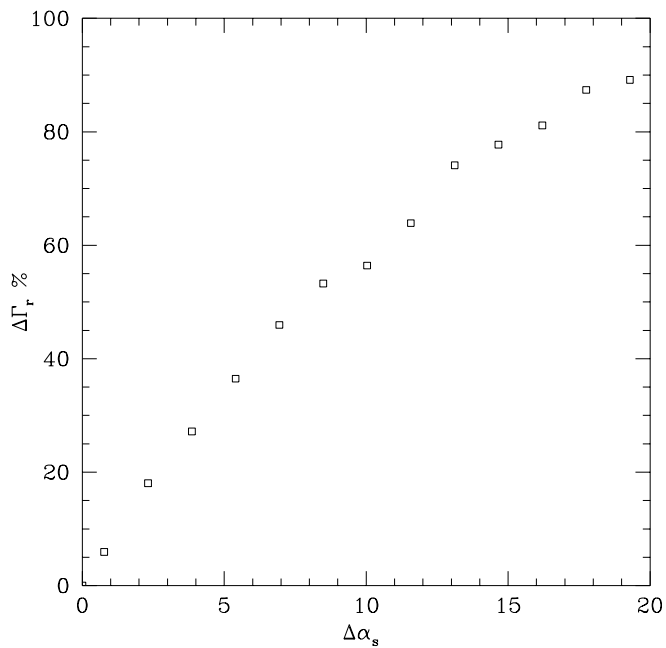,height=9.cm,width=9.cm}
\caption{The average error, expressed in percent,
  on the reconstructed value of $\Gamma$ as a function of the error of
  the orientation,$\alpha_s$,   of the  source  galaxy,  expressed  in
  degrees, One can see that if $\alpha_s$  is given with a better than
  $5^o$ precision then the error on $\Gamma$ is less than $30\%$.}
\label{errsph}
\end{figure}

\unitlength=1cm

\begin{center}
\begin{figure}
\begin{picture}(14,7)
\put(0.,0.){
\psfig{file=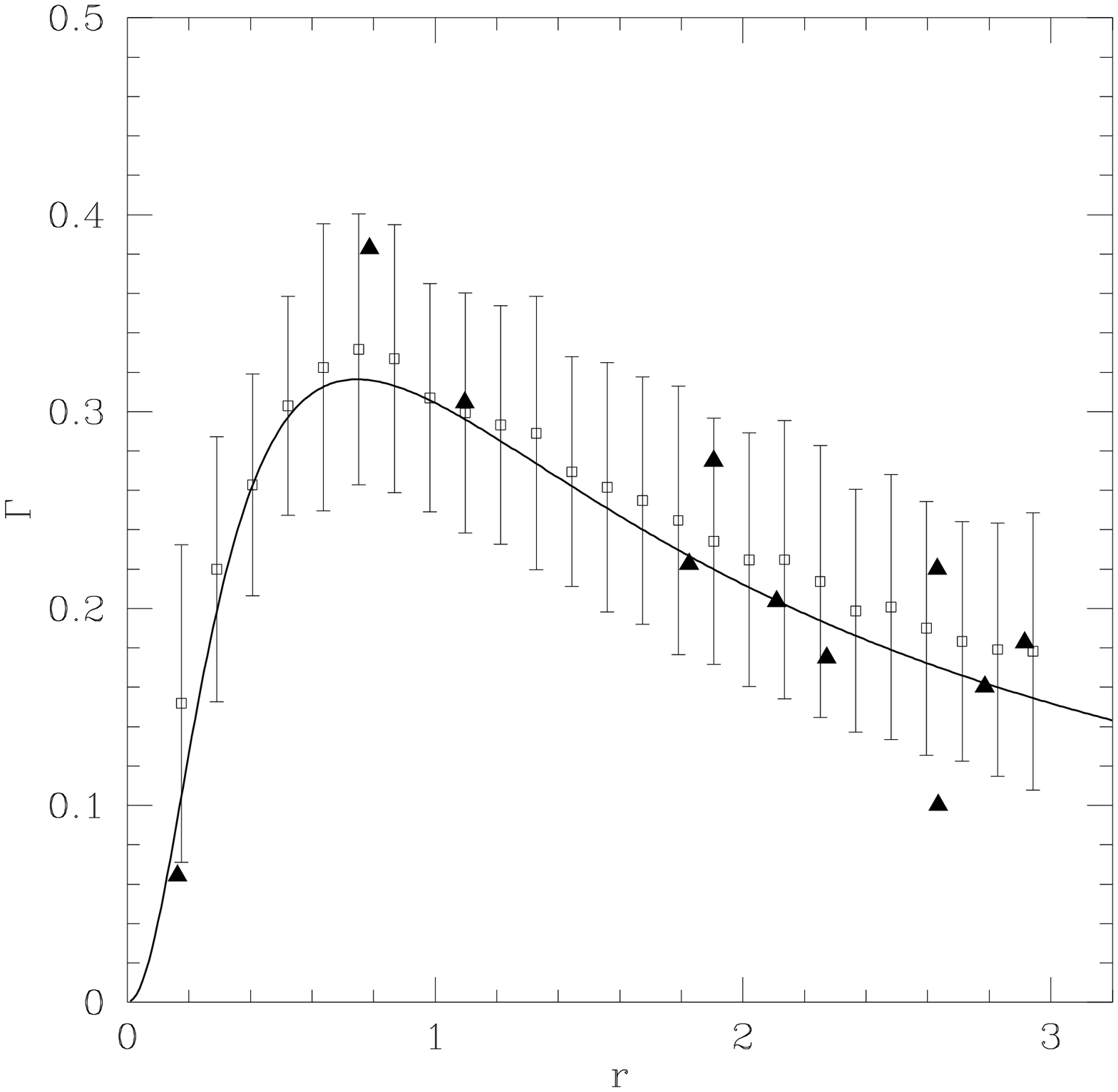,height=7cm,width=7cm}}
\put(7.0,0){
\psfig{file=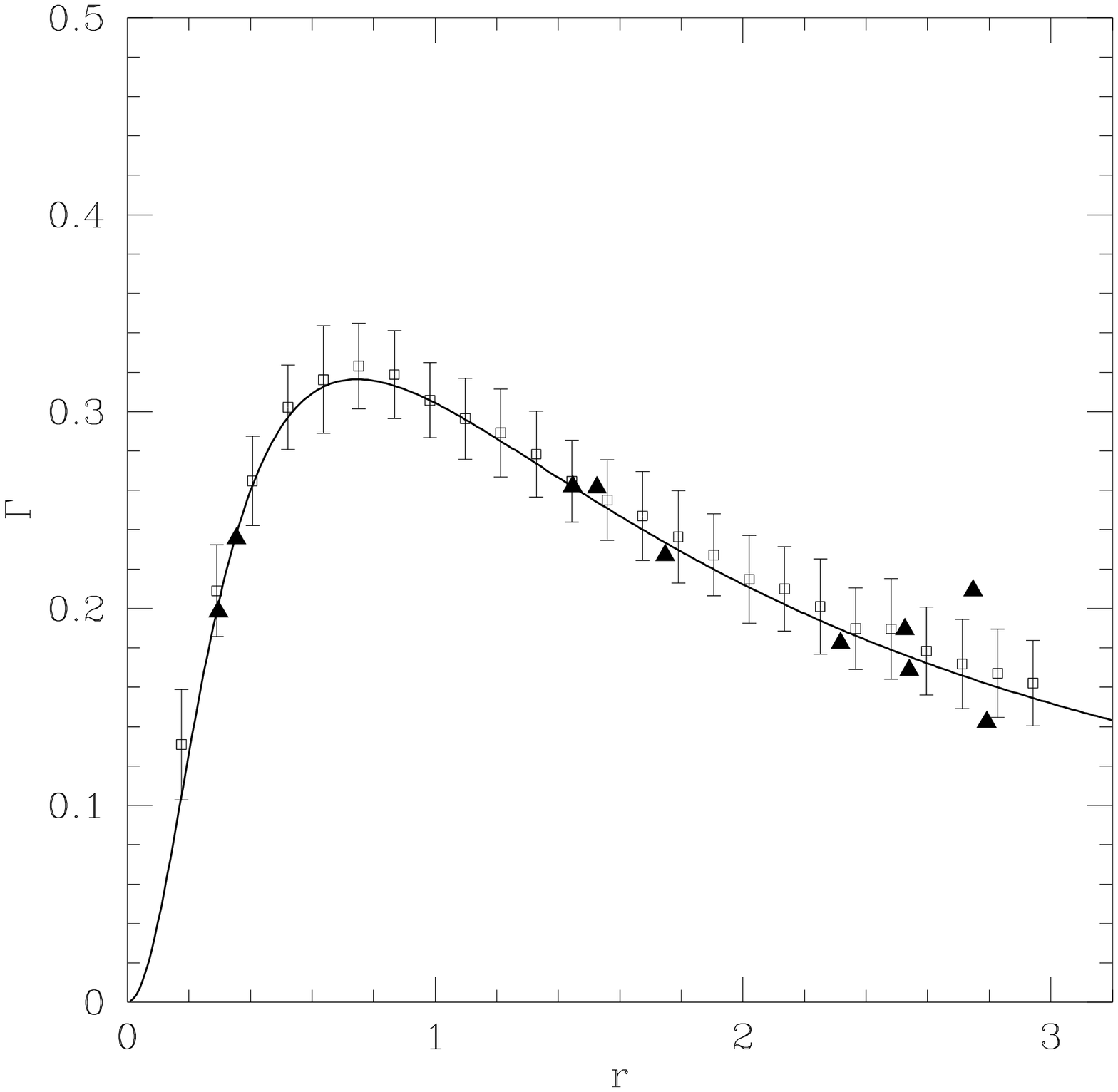,height=7cm,width=7cm}}
\end{picture}
\caption{This figure shows the value and the error (square) on the parameter
  $\Gamma$ as a  function of radius. The bars  are $95\%$ error bars.  
  The filled  in  triangles correspond to  the  values found for  $10$
  randomly selected galaxies.  The left (right) diagram corresponds to
  a population of  galaxies  with a $1-\sigma$  dispersion  of $2.5^o$
  ($5^o$) on the measure of $\alpha_s$. }
\label{recsph}
\end{figure}
\end{center}

%(Comme $\Gamma \in [0,1]$ cette  equation  permet de trouver  $\Gamma$
%sans ambiguité).

If there is no measurement error and if the  lens is really spherical,
this   method   would provide  an   exact  determination  of  the lens
parameters. Of course these assumptions are not very realistic, and it
is  interesting   to  investigate  the  case described    by  equation
(\ref{solr}) when there is a measurement error $\alpha_s$. In order to
carry out this  analysis, we have reconstructed  $\Gamma$ for $10 000$
randomly selected galaxies as before, and including  a random error on
the measurement of  $\alpha_s$.  Fig.\ref{errsph}  shows  the mean
error     in   percent   on    the   value     obtained for   $\Gamma$
($\Delta\Gamma_{r}=(\Gamma_{r}-\Gamma)/\Gamma$ where $\Gamma_{r}$  and
$\Gamma$ are  the inferred and the actual  value) as a function of the
error,         measured        in     degrees,       on     $\alpha_s$
($\Delta\alpha_{s}=|\alpha_{s}^{r}-\alpha_s|$ ). Evidently, if one can
measure $\alpha_{s}$ to   within  a few   degrees, it is  possible  to
locally determine $\Gamma$ to a precision of  around $20\%$ using just
one galaxy.

We  generated   two samples   of    galaxies with  a   gaussian  error
distribution on $\alpha_s$ with a  standard deviation of $2.5^{o}$ and
$5^{o}$.  The mean   value and the   one sigma  error  on the  on  the
retrieved  values $\Gamma$   are shown   in Fig \ref{recsph}. This
figure shows   that even with  only ten  or so galaxies  for which the
polarization hase  been measured, it is  possible to fairly accurately
determine the function $\Gamma(r)$. It is also  important to point out
that if one uses two  galaxies  at similar  radial distances from  the
centre, it is possible to test the assumption of spherical symmetry.

\subsection{Arbitrary lenses}

In   this section  we discuss   the  problem of  determining  the lens
parameters when no  assumptions  are made about  the  lens.  If we  no
longer assume that  the lens is  spherical, then the angle $\theta$ is
not known a  priori.  The system of  equations (\ref{sr1}),(\ref{sr2})
and   (\ref{sr4})   then consists  of   $3$  equations for  4 unknowns
($l_s,\Delta l_s, \Gamma,  \theta$), and so  it is impossible to solve
for the unknowns. However if we assume that  there are two galaxies in
the sample that can effectively be  associated with the same values of
$\Gamma$  and $\theta$,  then the  previous  system  written for  both
galaxies  will contain   $6$ unknowns,  $(l_{s}^{1},  l_s^{2},  \Delta
l_{s}^{1},   \Delta  l_{s}^{2}, \Gamma, \theta)$,   and six equations,
which means that it can  be solved exactly  for $\Gamma$, $\theta$ and
the source galaxies' parameters. One should note, however, that if the
orientation of the source galaxies  is not known, adding more galaxies
in this  way  cannot close the  system,  as there will always  be more
unknowns  than  independent equations: each    new galaxy brings three
unknowns and  three equations,  and   so   the equations are    always
underdetermined.  There  are a number of  methods for finding galaxies
associated with the  same pairs of values  of $(\Gamma, \theta )$.  If
we make no   assumptions about the   lens, then one  has to   take two
galaxies that have an angular separation much smaller than the angular
diameter of the lensing cluster.   One can then make the approximation
that the potential  is the same  for both galaxies.   If on the  other
hand  we assume that the  potential is spherically symmetric, then one
can take galaxies  which are well separated but  at  the same distance
from the centre,  so they have the  same $\Gamma$ and $\theta$.  There
could of course be other symmetries that allow one to do this.

The equation   (\ref{solr})  for each   galaxy gives  the  following
system:
\begin{equation}
\left\lbrace
\begin{array}{l}
(C_i^{g_1} S_s^{g_1} + S_i^{g_1} C_s^{g_1})\Delta l_{i}^{g_1}
\Gamma^{2} + 2 S_s^{g_1} \Gamma
+( C_i^{g_1} S_s^{g_1} - S_i^{g_1} C_s^{g_1})\Delta l_{i}^{g_1}  = 0 \\
(C_i^{g_2} S_s^{g_2} + S_i^{g_2} C_s^{g_2})\Delta l_{i}^{g_2}
 \Gamma^{2} + 2 S_s^{g_2} \Gamma
+( C_i^{g_2} S_s^{g_2} - S_i^{g_2} C_s^{g_2})\Delta l_{i}^{g_2}  = 0
\end{array}
\right.
\label{deuxgal}
\end{equation}
where $g_1$  (resp.  $g_2$)  denote  quantities related to  the  first
(resp. second) galaxy.  The parameters $C_{k}^{g_l}$ and $S_{k}^{g_l}$
($k=i,s$)  depend only  on $\theta$. This   system  can be  solved for
$\Gamma$   and $\theta$.  As before,   we   tested the possibility  of
reconstructing lens  parameters by  generating a  sample of   $10 000$
galaxies. Two    galaxies  are selected   which   correspond  to  lens
parameters which are only  slightly different. Thus we have considered
only pairs of galaxies separated by less than $0.2$ in the lens plane.
To simplify the presentation we shall use the following notation.

\begin{itemize}
\item $\Gamma^{g_i}$ parameter corresponding to the $i^{th}$ galaxy.
\item $\Gamma$ retrieved value
\item $\Gamma_m=(\Gamma^{g_1}+\Gamma^{g_2})/2$, the average value.
\item $\Delta\Gamma_{12}=|\Gamma^{g_1}-\Gamma^{g_2}|/\Gamma_m$:
the initial difference in $\Gamma$ for the two galaxies.
\item $\Delta\Gamma =|\Gamma-\Gamma_m|/\Gamma_m$: the error on the
retrieved value
\item  $\Delta\alpha_s=(\Delta\alpha_s^{g_1}+\Delta\alpha_s^{g_2})/2.$
 The quantities $\Delta\alpha_s^{g_k}$ are defined as in the paragraph above.
\end{itemize}

\begin{figure}
  \psfig{file=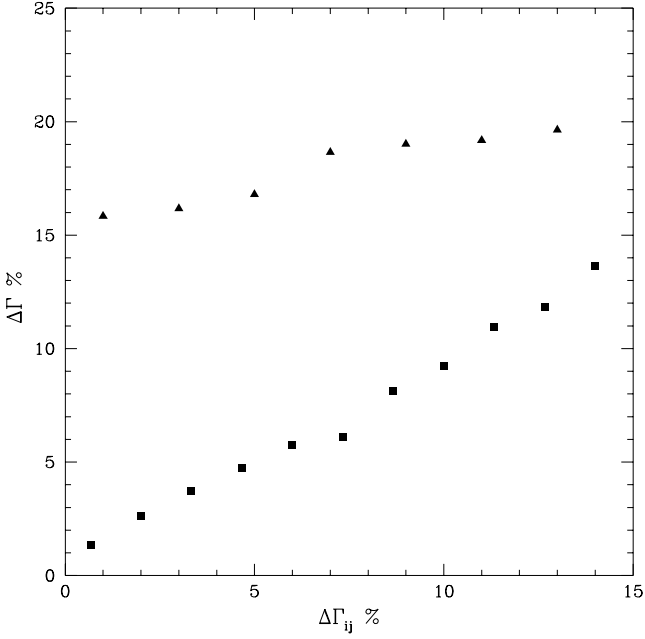,height=9.cm,width=9.cm}
\caption{Error on $\Gamma$ plotted against the difference between 
  $\Gamma^{g_1}$ and $\Gamma^{g_2}$ (see  text) for galaxies for which
  $\alpha_s$ is measured exactly (squares) and a standard deviation of
  $2.5^o$.}
\label{rpf1}
\end{figure}

In Fig.\ref{rpf1} we   have    plotted the average   value    of
$\Delta\Gamma$ as a function  of  $\Delta\Gamma_{12}$. If there  is no
measurement error, the  error   on the retrieved  $\Gamma$  is roughly
equal to the    difference   between the  values  of    $\Gamma_{g_1}$
and$\Gamma^{g_2}$.  If on the other hand  one  plotsthe same curve but
for  a population of  galaxies when a  measurement error of $2.5^o$ is
made,  then the error on   $\Gamma$ is noticeably  greater, and around
$15-20\%$. In order to establish the precision on $\alpha_s$ necessary
to establish  $\Gamma$  up to a  certain level  of  accuracy,  we have
plotted the mean  value of $\Delta\Gamma$ against $\Delta\alpha_s$ for
pairs of  galaxies  satisfying  $\Delta\Gamma_{12}<10\%$.   Indeed one
sees  that just one pair of  nearby galaxies whose polarization yields
$\alpha_s$ with $5^o$ accuracy  allows $\Gamma$ to be  determined with
an accuracy of $25\%$ without making any assumptions about the form of
the lens. Evidently, if  one has a large  number of galaxy pairs, that
it is possible to still further reduce this error.

One should also not  that  if one has   several pairs of  neighbouring
galaxies, the  number of equations in  (\ref{deuxgal}) is greater than
the number of unknowns, and  one should use statistical techniques for
the determination of the lens parameters.

\begin{figure}
\psfig{file=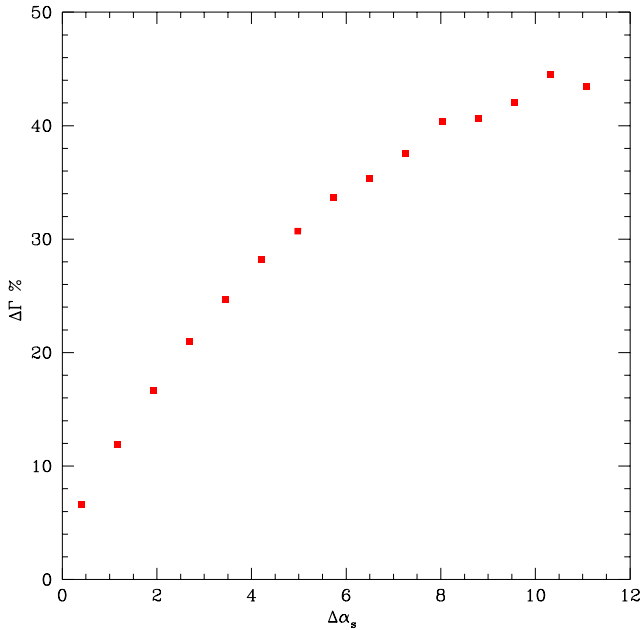,height=9.cm,width=9.cm}
\caption{Error on $\Gamma$ plotted against the average error in $\alpha_s$.}
\label{das}
\end{figure}

\section{Determination of $\kappa$ and $\gamma$}

Once $\Gamma$ has been determined, it is  still necessary to determine
the  convergence  $\kappa$  and  the  shear  $\gamma$,  which  are the
parameters  directly  linked to the mass  density  distribution in the
lens  plane.   These   parameters can     be obtained from    equation
(\ref{s2}), which can be written in the form

\begin{equation}
\lambda_{s} = \kappa^{2} \left( \lambda_{i} (1+\Gamma^2) + 2
\Delta\lambda_{i} \Gamma C_{i} \right) = a \kappa^{2} .
\end{equation}

If we now average the above relation over a  subset $S$ of galaxies
having   the same value  of  $\kappa$,  either because they are
sufficiently close together or  because of some symmetry (all the
galaxies can be used and not only those with a measured
polarization)we obtain $\kappa^2 = \bar{\lambda_{s}}/ \bar{a}\, ,$
where  $\bar{a}$,   the average of  $a$ over   the  subset $S$, can be
computed using the  image and the  reconstructed value of $\Gamma$ and
$\theta$ (which are  equal for all galaxies  in $S$).  The  average of
$\lambda_{s}$,  $\bar{\lambda_{s}}$,  cannot  be  computed  since  the
intrinsic  brightness of  the sources   are not  known,  but since the
source  galaxies   (and  therefore  their  properties)  are  uniformly
distributed   in   the source   plane,  $\bar{\lambda_{s}}$  should be
constant (independent of the set  $S$ chosen) if  the average is taken
over a  sufficiently  large   number of   galaxies. Therefore we   can
arbitrarily  set  $\bar{\lambda_{s}}$=1  for    the whole   subset  of
galaxies.  This allow  to   compute  $\tilde{\kappa}=1/  \bar{a}$  and
$\tilde{\gamma}=\Gamma \tilde{\kappa}$  which    are   proportional to
$\kappa$ and $\gamma$ respectively, the proportionality factor being a
constant throughout  the lens plane.  The  reliability  of this method
will strongly depend on the  number of galaxy  contained in each subset
and  on the luminosity dispersion  of the source galaxies. Each subset
should contain   enough  galaxies    in  order for   to    assume  the
$\bar{\lambda_{s}}$ is equal to the  average luminosity of the  source
galaxies.

\section{Observational considerations}

In the previous section we saw how knowledge of the orientation of the
source  galaxy    enabled  us to improve    the   determination of the
gravitational field  of  the cluster.   In principle  this orientation
would  be  given  by the  direction   of  the polarization.  Similarly
measurement of the degree  of polarization would yield  information about
the  ellipticity of  the source,  although   in view  of the  possible
uncertainty concerning the  mechanisms  involved in  the production of  the
polarization   detailed modelling might   be   required.  Potentially,
determination  of the intrinsic polarization  of lensed galaxies could
yield valuable information. Thus  a question of crucial  importance is
the feasibility of such measurements and the reliability of their
interpretation.

The main difficulty evidently arises  from the faintness of the source
galaxy, and the  relatively low level  of polarization.   It should be
borne in mind,  however, that  although we are  looking at  objects of
magnitude 25 or fainter, it is the integrated polarization that we are
interested in, and measurements could  be taken over broad band.  Such
observations in the optical have been carried out for a number of high
redshift radio  galaxies  (Tadhunter et al.   1992), with the galaxies
showing high  polarization,  which appears   to  increase with  higher
redshift.  Although our concern is mainly  with spiral galaxies, it is
possible that radio galaxies could also be used in the analysis, since
these   appear  to display net  polarization  oriented   in a definite
direction  to  the radio axis.  Our main  point  here  however is that
optical polarimetric  measurements   would   be  difficult, but    not
impossible, particularly  on a  large  telescope.  It is  important to
point out also   that in our  analysis  it  is only  the  direction of
polarization that  is important, and  this is much easier to establish
accurately than the degree of polarization (Treanor 1968).

Another problem would be the  possible effect of Faraday rotation, yet
in the visible we would expect this to  be fairly weak.  If we were to
use polarimetric measurements in the radio, then  this of course would
need to be taken into account.

\section{conclusions}

The polarimetric  study of  weakly  lensed galaxies can   in principle
provide information  about  the lensing gravitational  potential  that
otherwise would be  extremely difficult or  impossible to obtain.  The
method relies on the property that the  plane of polarization of light
is not affected by the  gravitational potential. Thus the direction of
polarization  provides  a marker   for the  orientation of the  source
galaxy, and  hence   information about the  distortion    in the image
induced by the lens. This reduces the uncertainty in the estimation of
the lens  parameters, and can even,  in certain situations  when a few
source  galaxies can actually  be measured polarimetrically, break the
degeneracy of the equations that determine the lens parameters.

Although to a  large extent we  have assumed that Thomson  or Rayleigh
scattering  is   responsible  for the    polarization   flux, the  our
conclusions  are not greatly  affected by  the presence  of such small
scale  mechanisms as dichroism provided that  on the large scale axial
symmetry holds.  Only  in exceptional  circumstances would one  expect
the direction of the polarization not to be perpendicular to the major
axis of the  elliptical isophotes.  Since  weak lensing would induce a
rotation   of much  less than   90  degrees,  even if the   integrated
polarization were aligned parallel to  the major  axis of the  source,
this  should   not lead  to confusion  and   hence measurement  of the
direction of polarization would  still provide rigorous constraints on
the lensing parameters.

Measurement  of  the optical   polarization of  faint   galaxies would
require large telescopes and integration times, but it is important to
realise that  the main advantage of  the proposed method relies on the
use of   the  direction  of   polarization, and  not  the  degree   of
polarization and the former is considerably easier to establish.

\end{document}